\documentclass[
twocolumn,
prl,
showpacs,
amsmath,
amssymb,
superscriptaddress,
groupedaddress,
floatfix
]{revtex4}
\usepackage{bm}
\usepackage{dsfont}
\usepackage{graphicx}
\usepackage{pifont}
\usepackage{textcomp}
\usepackage{gensymb}
\usepackage{accents}
\usepackage{upgreek}
\usepackage{color}

\newcommand{\s}{\sum\limits}

\newcommand{\be}{\begin{equation}}
\newcommand{\e}{\end{equation}}
\newcommand{\bbp}{\begin{minipage}}
\newcommand{\emp}{\end{minipage}}
\newcommand{\beml}{\begin{subequations}}
\newcommand{\epm}{\end{pmatrix}}
\newcommand{\bpm}{\begin{pmatrix}}
\newcommand{\bc}{\begin{cases}}
\newcommand{\ec}{\end{cases}}
\newcommand{\eml}{\end{subequations}}
\newcommand{\beq}{\begin{eqnarray}}
\newcommand{\eq}{\end{eqnarray}}
\newcommand{\ba}{\begin{array}}
\newcommand{\ea}{\end{array}}
\newcommand{\lt}{\left}
\newcommand{\rt}{\right}
\newcommand{\n}{\nonumber}

\newcommand{\ep}{\varepsilon}

\newcommand{\bb}{\boldsymbol}

\newcommand{\h}{^\dagger}

\makeatletter
\def\NAT@bibsetnum#1{%
 \setlength{\topsep}{\z@}%
 \NATx@bibsetnum{#1}%
}%
\renewenvironment{thebibliography}[1]{%
 \NAT@thebibliography{#1}%
 \@clubpenalty\clubpenalty
 \let\@TBN@opr\present@bibnote
 \@FMN@list
}{%
 \@endnotesinbib
 \edef\@currentlabel{\arabic{NAT@ctr}}%
 \NAT@endthebibliography
 \global\let\auto@bib\@empty
}
\newcommand*{\supplementarystart}{%
  \close@column@grid%
  \clearpage%
  \onecolumngrid%
  \setcounter{enumiv}{0} 
  \setcounter{equation}{0} 
  \setcounter{figure}{0} 
  \setcounter{table}{0} 
  \setcounter{page}{1}
  \c@secnumdepth=4
  \renewcommand{\theequation}{s\arabic{equation}} 
  \renewcommand{\bibnumfmt}[1]{[s##1]} 
  \renewcommand{\@onlinecite}{s\citealp} 
  \renewcommand{\cite}[1]{{[}\onlinecite{##1}{]}}
  \renewcommand{\thefigure}{s\arabic{figure}}
  \renewcommand{\thetable}{s\Roman{table}}
  \renewcommand{\thepage}{s\arabic{page}}
}
\makeatother

\begin{document}

\title{Light-induced anisotropic skyrmion and stripe phases in a Rashba ferromagnet}

\author{Dmitry Yudin}
\affiliation{ITMO University, Saint Petersburg 197101, Russia}

\author{Dmitry R. Gulevich}
\affiliation{ITMO University, Saint Petersburg 197101, Russia}

\author{Mikhail Titov}
\affiliation{Radboud University Nijmegen, Institute for Molecules and Materials, NL-6525 AJ Nijmegen, The Netherlands}
\affiliation{ITMO University, Saint Petersburg 197101, Russia}

\date{\today}

\begin{abstract}
An external off-resonant pumping is proposed as a tool to control the Dzyaloshinskii-Moriya interaction (DMI) in ferromagnetic layers with strong spin-orbit coupling. Combining theoretical analysis with numerical simulations for an $s$-$d$--like model we demonstrate that linearly polarized off-resonant light may help stabilize novel non-collinear magnetic phases by inducing a strong anisotropy of the DMI. We also investigate how with the application of electromagnetic pumping one can control the stability, shape, and size of individual Skyrmions to make them suitable for potential applications. 
\end{abstract}


\maketitle

Low-dimensional magnetic structures provide an exciting playground for condensed matter physics and technology applications. Some of them, e.\,g., helical magnets \cite{Ishikawa1976,Moriya1982,Pfleiderer2004,Li2012}, are known to support topologically nontrivial magnetic textures \cite{Kishine2011,Togawa2012,Koumpouras2016}. Such noncollinear states emerge as a result of the competition between Heisenberg exchange, antisymmetric Dzyaloshinskii-Moriya exchange, and magnetocrystalline anisotropy yielding magnetic ground states that are far more intricate than those in homogeneous ferromagnets \cite{Bak1980,Maleyev2006,Belitz2006,Vedmedenko2007,Grigoriev2009}. Recent progress in the fabrication of magnetic materials motivated an interest in particlelike domains, such as magnetic Skyrmions \cite{Muhlbauer2009,Yu2010,Heinze2011,Yu2011,Kiselev2011,Tonomura2012,Nagaosa2013,Rybakov2013,Pereiro2014,Gayles2015,Keesman2015,Gungordu2016,Leonov2016b}, that are typical for nonhomogeneous ferromagnets. Skyrmions and other topologically protected magnetic textures have been proposed as building blocks for logical operations and information storage in the rapidly advancing fields of magnon spintronics \cite{Chumak2015} and Skyrmionics \cite{Krause2016,Leonov2016a}. 

Spintronics, as a branch of applied science, is traced back to the pioneering work on giant magnetoresistance by Gr\"unberg \cite{Saurenbach1988,Binasch1989} and Fert \cite{Baibich1988}. The subsequent discovery of spin-transfer torque \cite{Slonczewski1989,Berger1996} provoked the idea to exploit the spin degree of freedom rather than the charge for processing and transferring information. This idea is currently being extended to magnetic materials supporting localized magnetic excitations such as Skyrmions. 

Ways to create and control Skyrmions and other noncollinear magnetic textures are essential for the practical implementation of this emerging technology. In this Letter, we investigate microscopically how the noncollinear magnetic domains in thin ferromagnet layers with strong spin-orbit coupling may be controlled by linearly polarized light. The effects predicted may be observed in thin films such as Co/Pt heterostructures subject to short light pulses \cite{Huisman2016}.

The exchange interaction alone may lead only to a collinear orientation of magnetic moments in a cubic crystal. Spatially inhomogeneous magnets are usually associated with a lack of lattice inversion symmetry. In his seminal work on noncentrosymmetric magnets \cite{Dzyaloshinskii1964}, Dzyaloshinskii identified the one-dimensional magnetic spiral states stabilized by the Dzyaloshinskii-Moriya interaction (DMI) \cite{Dzyaloshinskii1958} that favors the noncollinear orientation of neighboring spins. A nontrivial ground state arises in helical magnets as a consequence of the competition between the Heisenberg exchange and the DMI. In two-dimensional structures, this competition leads to a helical spin-spiral ground state configuration that becomes unstable in the presence of a magnetic field with the tendency towards the formation of Skyrmions \cite{Rosler2006}. Quite generally, Skyrmions correspond to the solitonlike solutions of the field equations of Dzyaloshinskii's theory that destroy the homogeneity of magnetic order \cite{Bogdanov1989}. The existence of such localized states and the mechanism of their nucleation as mesoscopic objects are rather common for continuum systems described by the free energy functional with Lifshitz invariants \cite{Rosler2011}. The strength of DMI can be rigorously approached by the microscopic theory, as an indirect exchange interaction between two neighboring spins facilitated by itinerant (conduction) electrons \cite{Imamura2004,Kundu2015,Kikuchi2016}, as well as adopted from the first-principles simulations \cite{Nossa2012,Dmitrienko2014,Koretsune2015}. In a thin film, the DMI can be induced by the Rashba spin-orbit coupling.

An external electromagnetic field may strongly modify the properties of an electronic system providing an important tool for manipulating materials in a controllable fashion \cite{Mikhaylovskiy2015,Fujita2017a,Fujita2017b}. 
It has been recently shown that the effect of off-resonant electromagnetic radiation (with the frequency exceeding the bandwidth of the system) may be described by effective time-independent models with strongly renormalized parameters \cite{Oka2009,Kitagawa2011,Lindner2011,Wang2013,Cayssol2013,Yudin2016a,Yudin2016b,Stepanov2016,Yudin2017,Sato2017}. 

In this Letter, we derive the effective $s$-$d$ exchange model of a Rashba ferromagnet in the regime of strong coupling to external radiation. We find that the DMI strength can be effectively controlled by the application of off-resonant pumping that opens up exciting opportunities for controlling the stability, size, and shape of individual metastable Skyrmions. Also, we show that the application of linearly polarized radiation induces anisotropy of the DMI that not only provides a finer control over the individual DMI strengths in two orthogonal directions but also leads to the appearance of novel anisotropic phases. 

From the theory point of view, the effect of time-periodic fields may be described, with some reservations, by the so-called Floquet theory \cite{Grifoni1998,Kohler2005}.
Periodicity of the driving field enables one to map the original time-dependent problem to the eigenvalue problem of Floquet states. Off-resonant pumping takes place if the frequency of the driving field is so high that electrons are not able to follow field oscillations. In this case, real absorption or emission of light quanta cannot happen due to restrictions imposed by energy conservation for radiation frequency exceeding the electron bandwidth. Still, such off-resonant radiation affects the system via virtual processes leading to a significant renormalization of the parameters of the initial Hamiltonian of an electron subsystem. Below, we restrict our attention to the effects of linearly polarized light, since it has a greater impact on the noncollinear magnetic textures. For the sake of microscopic treatment, we rely upon the Floquet-Magnus expansion and its generalizations that have been developed in Refs.~\cite{Blanes2009,Goldman2014,Eckardt2015,Bukov2015}.

For microscopic analysis, we consider a weak two-dimensional ferromagnet that yields an $s$-$d$-like Rashba model for conduction electrons: 
\be
\label{hamiltonian}
H=p^2/2m+\alpha\, \bb{\sigma}\!\cdot\!\lt(\bb{p}\times \hat{\bb{z}}\rt)+\Delta\,\bb{m}\!\cdot\!\bb{\sigma},
\e
where $\bb{m}$ is the unit ($|\bb{m}|=1$) local magnetization vector due to, e.\,g., localized $d$ electrons, $\Delta$ is the $s$-$d$-like exchange energy, $\bb{p}$ stands for the momentum operator for conduction electrons with an effective mass $m$, $\alpha$ is the Rashba spin-orbit interaction strength, $\hat{\bb{z}}$ is the unit vector in the direction perpendicular to the two-dimensional electron gas, and $\bb{\sigma}=\lt(\sigma_x,\sigma_y,\sigma_z\rt)$ denotes the vector of Pauli matrices.  Models of the type of Eq.~(\ref{hamiltonian}) were originally proposed to explain the physics of ferromagnetic metals beyond the Heisenberg exchange picture \cite{Vonsovsky1946,Zener1951a,Zener1951b,Zener1951c}. This approach relies upon a formal distinction between a localized (classical) magnetic subsystem (e.\,g., $d$ or $f$ electrons, that are described by an $\bb{m}$ field which is governed by a classical Heisenberg model) an itinerant subsystem [e.\,g., $s$ electrons described by Eq.~(\ref{hamiltonian})] that are coupled to each other by means of exchange interaction. 

An external {\em ac} electromagnetic field of frequency $\omega$ \cite{footnote} is introduced in the model of Eq.~(\ref{hamiltonian}) by means of the Peierls substitution $\bb{p}\to\bb{p}+e\bb{A}_0\cos\omega t$, where $\bb{A}_0=\bb{E}_0/\omega$ and $\bb{E}_0$ is the electric field component of the field.  In what follows, we restrict our analysis to the case of linearly polarized light by choosing $\bb{E}_0=E_0\hat{\bb{y}}$, where $\hat{\bb{y}}$ is the in-plane unit vector (in the $y$ direction). 

\begin{figure}
\centerline{\includegraphics[width=0.9\columnwidth]{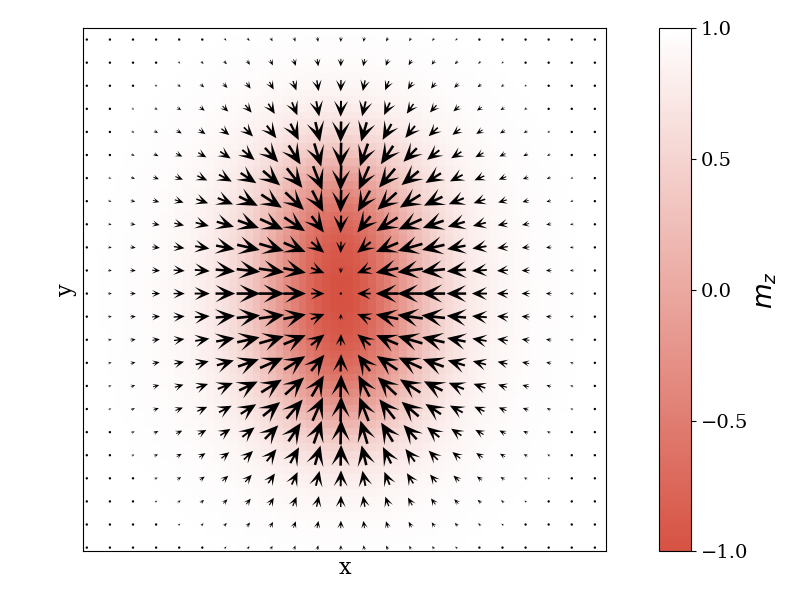}}
\caption{
Elliptic Skyrmion arising in the presence of the DMI anisotropy induced by the linearly polarized light. The figure represents the results of our numerically simulations for the model of Eq.~(\ref{E}) at $H_\mathrm{ext}=0.01 J$ and $\gamma=1.5$ for the DMI strength fixed at $D_x=D_y=0.18 J$ for $\gamma=0$. Arrows indicate the in-plane components of the average magnetization in a block of $3\times 3$ spins.}
\label{fig:skyrmion} 
\end{figure}

The Hamiltonian of Eq.~(\ref{hamiltonian}) is conveniently rotated as $H\to U_t\h H U_t$ with the time-dependent unitary operator
\beml
\begin{align}\label{matrix1}
&U_t =2^{-1/2}(\sigma_x+\sigma_z)e^{-i t \lt[u_t+\sigma_z\,e\alpha A_0\sin(\omega t) /\omega t\rt]/\hbar},\\ \label{matrix2}
&u_t =\frac{e^2A_0^2}{4m}\lt(1+\frac{\sin(2\omega t)}{2\omega t}\rt)+\frac{ep_yA_0}{m}\frac{\sin(\omega t)}{\omega t}.
\end{align}
\eml
The transformed Hamiltonian yields the matrix Floquet model of the form
\be
\label{hamiltonian2}
H=\frac{p^2}{2m}+\lt(\alpha p_y+\Delta m_x\rt)\sigma_z+ \sum_{n=-\infty}^\infty \!\!\!h_n\, e^{i n\omega t},
\e
with the coefficients $h_n$ defined by 
\be
h_n=\lt[\Delta m_z\sigma_x+\lt(\alpha p_x-\Delta m_y\rt)\sigma_y\rt]J_n(\gamma),
\e
where the parameter $\gamma=2e\alpha E_0/\hbar\omega^2$ describes the effective light-matter coupling and $J_n(\gamma)$ stands for the $n$th order Bessel function of the first kind.
 
The high-frequency expansion in the form of the Brillouin-Wigner perturbation theory recently developed for this class of problems \cite{Mikami2016} maps Eq.~(\ref{hamiltonian2}) onto an effective time-independent Hamiltonian:
\beml
\label{model}
\begin{align}
\label{modelA}
&H_\mathrm{eff}=p^2/2m+\alpha\lt[p_y \sigma_x- p_x \sigma_y\,J_0(\gamma)\rt]+V,\\
&V=\Delta \{m_x\sigma_x+\lt[ m_y\sigma_y+m_z\sigma_z\rt]J_0(\gamma)\},
\end{align}
\eml
that is valid away from resonance frequencies. The effective model fails only in a tiny vicinity $\delta\gamma$ of the zeros of the Bessel function, $\delta\gamma\approx 10^{-5}\Delta^2/(\hbar\omega)^2$, which is well beyond our numerical resolution \cite{sup}. The model is equivalent to that of Eq.~(\ref{hamiltonian}) with anisotropic renormalization of coupling constants: Rashba spin-orbit interaction strength and $s$-$d$ exchange coupling. In what follows, we assume a weak ferromagnet and treat the exchange interaction term $V$ perturbatively. 

Based on the symmetry analysis, Dzyaloshinskii discovered that the effective Ginzburg-Landau free energy functional may allow for terms linear in magnetization gradients provided the absence of lattice inversion symmetry \cite{Dzyaloshinskii1958}. Later, Moriya argued, on the basis of Anderson's theory of superexchange, that the microscopic mechanism of spin-orbit coupling is responsible for such an interaction \cite{Moriya1960a,Moriya1960b}. The latter can also be  thought as a coupling between an excited state of a magnetic ion and the ground state of the neighboring ion. Such a coupling can be derived microscopically from the correction to the bare action $\mathcal{S}_0[\bb{m}]$ (that collects all terms corresponding to magnetic subsystem) computed to the second order with respect to the perturbation $V$ \cite{sup}. 

\begin{figure*}
\centerline{
\includegraphics[width=2\columnwidth]{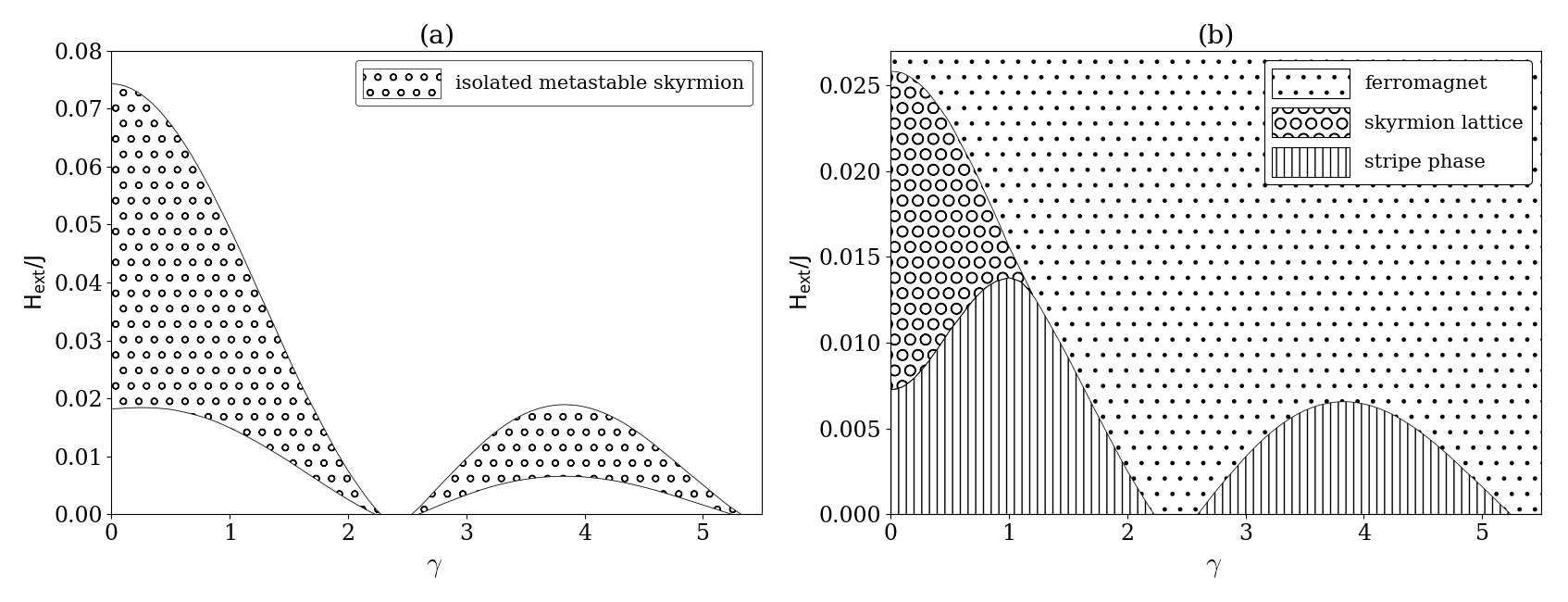}
}
\caption{
(a) Region of metastability of an isolated Skyrmion in the model of Eq.~(\ref{E}) with the anisotropic DMI strength given by Eq.~(\ref{dzyamo}). (b) Phase diagram as a function of the parameter $\gamma$ and external magnetic field $H_\mathrm{ext}$. Both figures represent the results of numerical simulations of the model of Eq.~(\ref{E}) with the DMI strength given by $D_x=D_y=0.18 J$ at $\gamma=0$.
}
\label{fig:phd} 
\end{figure*}

To construct the perturbation theory, we take advantage of the bare Matsubara Green's function for the Hamiltonian of Eq.~(\ref{modelA}): 
\beml
\begin{align}
&G_0(i\omega,\bb{k})=\frac{1}{2}\s_{s=\pm}\frac{\Lambda_s(\theta_{\bb{k}})}{i\hbar\omega-\ep^s_{\bb{k}}},\\
&\Lambda_\pm(\theta)=1\pm\lt[\sigma_x\sin\theta-\sigma_yJ_0(\gamma)\cos\theta\rt]/g(\theta).
\end{align}
\eml
where $g(\theta)=\lt[\sin^2\theta+J_0^2(\gamma)\cos^2\theta\rt]^{1/2}$ and the spectrum $\ep_{\bb{k}}^\pm=\hbar^2k^2/2m \pm \alpha\hbar k\, g(\theta_{\bb{k}})$ acquires the dependence on the direction of the wave vector $\bb{k}=\bb{p}/\hbar=k (\cos\theta_{\bb{k}},\sin \theta_{\bb{k}})$ due to the linear polarization of light. 

The second-order contribution to the effective action in the imaginary time representation is given by
\be
\delta\mathcal{S}[\bb{m}]=\frac{1}{\beta}\s_{i\omega,\bb{k}} \s_{ij}\Pi_{ij}(i\omega,\bb{k})m_i(\bb{k})m_j(-\bb{k}),
\e
where $\beta$ stands for the inverse temperature while the indexes $i$ and $j$ denote the Cartesian vector components. The polarization operator is expressed as 
\begin{align}
\Pi_{ij}(i\omega,\bb{k})=\frac{1}{4}\Delta_i \Delta_j\s_{\bb{q}}\s_{s,s'=\pm}
\frac{f(\ep_{\bb{k}+\bb{q}}^s)-f(\ep_{\bb{q}}^{s'})}{\ep_{\bb{k}+\bb{q}}^s-\ep_{\bb{q}}^{s'}-i\hbar\omega}&\n\\ 
\times\mathrm{Tr}\lt[\Lambda_s(\theta_{\bb{k}+\bb{q}})\sigma_i\Lambda_{s'}(\theta_{\bb{q}})\sigma_j\rt],&
\label{polar}
\end{align}
where $f(\ep)$ is the Fermi-Dirac distribution function and $\Delta_x=\Delta$, $\Delta_y=\Delta_z=J_0(\gamma)\Delta$.

The straightforward expansion of $\Pi_{ij}(\omega=0,\bb{k})$ around $\bb{k}=0$ up to the terms linear in $\bb{k}$ yields a fully antisymmetric contribution to the effective action \cite{sup}: 
\be
\label{deltaS}
\delta\mathcal{S}[\bb{m}] = \int d^2r\lt(D_xL_{xz}^{(x)}+D_yL_{yz}^{(y)}\rt),
\e
where we introduce the DMI couplings
\be\label{dzyamo}
D_x=\frac{\Delta^2}{\pi\alpha\hbar}\frac{\lt|J_0(\gamma)\rt|}{1+\lt|J_0(\gamma)\rt|}, \quad 
D_y=\frac{\Delta^2}{\pi\alpha\hbar}\frac{J_0^2(\gamma)}{1+\lt|J_0(\gamma)\rt|},
\e
and Lifshitz invariants $L_{ij}^{(l)}=m_i\partial_l m_j- m_j\partial_l m_i$.

In the absence of an electromagnetic field, i.\,e., for $\gamma=0$, we obtain an isotropic DMI with $D_x=D_y=\Delta^2/(2\pi\alpha\hbar)$. In the presence of linearly polarized light, the DMI coupling becomes essentially anisotropic as given by Eq.~(\ref{dzyamo}). We stress that the employed high-frequency expansion is legitimate only as far as there are no resonant transitions and the parameter $\gamma$ is away from zeros of the Bessel function $J_0(\gamma)$ \cite{sup}. 

To illustrate our results, we consider a classical two-dimensional Heisenberg exchange model on a square lattice, that is given by the total energy
\begin{align}\n
&E=-J\s_{\bb{r}}\bb{S}_{\bb{r}}\cdot\lt[\bb{S}_{\bb{r}+\hat{\bb{x}}}+\bb{S}_{\bb{r}+\hat{\bb{y}}}\rt]-H_\textrm{ext}\s_{\bb{r}}\lt(\bb{S}_{\bb{r}}\rt)_z \\
&-D_x\s_{\bb{r}}\lt(\bb{S}_{\bb{r}}\times\bb{S}_{\bb{r}+\hat{\bb{x}}}\rt)_y+D_y\s_{\bb{r}}\lt(\bb{S}_{\bb{r}}\times\bb{S}_{\bb{r}+\hat{\bb{y}}}\rt)_x,
\label{E}
\end{align}
where $\bb{S}_{\bb{r}}$ is the spin on a lattice site $\bb{r}$, $H_\textrm{ext}$ is an external magnetic field (in energy units) perpendicular to the two-dimensional plane, $\hat{\bb{x}}$ and $\hat{\bb{y}}$ stand for the unit vectors in $x$ and $y$ direction, correspondingly, and the lattice constant is set to unity. 


With the help of the numerical approach described in Ref.~\cite{sup}, we analyze the influence of the anisotropic DMI on the Skyrmion profile. The profile is obtained by relaxing a trial Skyrmion ansatz using the dynamical Landau-Lifshitz-Gilbert equation until the stationary state is reached. To avoid nonuniversal effects of boundary conditions, the numerical simulation is performed in a box of a large size that exceeds characteristic size of a Skyrmion by a large factor (only the central part of the box is shown in Fig.~\ref{fig:skyrmion}).  

We find that the anisotropic renormalization of the DMI strength of Eq.~(\ref{dzyamo}) results in the anisotropic squeezing of a Skyrmion. The Skyrmion becomes elongated along the light polarization direction and develops an elliptic profile as shown in Fig.~\ref{fig:skyrmion}. In the model with $\gamma\lesssim 1$ and positive $D_x$ and $D_y$, the in-plane spin projections are directed towards the Skyrmion center. Such a configuration may be referred to as the inverted hedgehog Skyrmion that is distinguished from the hedgehog Skyrmion in which the in-plane spin projections point outwards. The Skyrmion type is, therefore, defined by the overall sign of the DMI coupling. 

It is also worth stressing that the model~\eqref{E} supports only Ne{\'e}l-type Skyrmions which are the Skyrmions with a radial orientation of spins. Such Ne{\'e}l-type Skyrmions were observed recently in GaV$_4$S$_8$~\cite{Kezsmarki2015}. Those should be contrasted with Bloch-type Skyrmions that are characterized by spin orientations perpendicular to the radial direction. The Bloch-type Skyrmions are thought to be characteristic for materials like FeGe \cite{Wilhelm2011} and MnSi \cite{Yu2015}. 


It can be shown using the methodology of Ref.~\cite{sup} that individual Skyrmions are metastable only in certain areas of the parameter space as illustrated in Fig.~\ref{fig:phd}(a). The metastability regions of individual Skyrmions in the model of Eq.~(\ref{E}) must be distinguished from the phases that characterize the absolute minimum of the energy functional. By extending our numerical analysis to search for a ground state \cite{sup}, we obtain the phase diagram depicted in Fig.~\ref{fig:phd}(b). The diagram consists of three phases: (i) the homogeneous ferromagnetic order phase denoted by points, (ii) the Skyrmion crystal phase (a crystal of elliptic Skyrmons) denoted by circles, and (iii) the stripe crystal phase (periodic stripes in the direction of light polarization) denoted by vertical lines.

Anisotropy induced by pumping distorts the symmetry of the Skyrmion crystal from the equilateral triangular at $\gamma=0$ to an isosceles triangular at nonzero $\gamma$ \cite{sup}. The stripe crystal phase is analogous to the helical phase discussed, e.\,g., in Ref.~\cite{Iwasaki-NComm-2013}, although, in contrast to the conventional helical phases, the stripe phase arising for the model~\eqref{E} is the N{\'e}el type, with spins rotating in the radial direction, as opposed to a helix. The orientation of the stripe phase depends on the direction of the induced anisotropy of the DMI. This provides a control over the orientation of the stripe phase by changing the polarization of the applied radiation --- a property which may be employed in future light-controlled magnetic logic gates.

Interestingly, the range of metastability of individual Skyrmions does not generally coincide with the phase boundaries. However, we find that metastable Skyrmions generally do not exist in a stripe crystal phase that is dominant at low values of the magnetic field. In this region, individual Skyrmions quickly become unstable with respect to stretching in the $y$ direction to form a stripe. For intense light with the parameter $\gamma$ exceeding the first zero of $J_0(\gamma)$, i.\,e., for $\gamma \gtrsim 2.4$, the phase diagram is dominated by the stripe phase at small magnetic fields. The Skyrmion crystal phase is limited to a moderate light intensity as shown in Fig.~\ref{fig:phd}(b).

The numerical studies of Skyrmion dynamics in the absence of the field $\gamma=0$ were performed in Refs.~\cite{Iwasaki-NN-2013, Iwasaki-NComm-2013, Iwasaki-Nano-2014, Koshibae-2016, ZZ-2016}. In the absence of light, i.\,e., for $\gamma=0$, the phase diagram of Fig.~\ref{fig:phd}(b) reproduces these known results. Indeed, the obtained values of the critical fields for the transition between the stripe and Skyrmion-crystal phases ($H_{c1}=0.0072 J$) and between the Skyrmion-crystal and ferromagnetic phases ($H_{c2}=0.026 J$) are very close to those given in Ref.~\cite{Iwasaki-NN-2013}. To simplify the comparison with Ref.~\cite{Iwasaki-NN-2013}, we have used the same parameter values of the DMI strength $D_x=D_y=0.18 J$ at $\gamma=0$.


In conclusion, the field of magnetic Skyrmions has attracted considerable attention due to the potential applications of Skyrmions in information processing. The major advantage of such noncollinear spin configurations as compared to domain walls is the possibility to make the Skyrmion size as small as a few nanometers without losing its stability. In this Letter, we employ the $s$-$d$-like exchange model for a weak two-dimensional ferromagnet with strong spin-orbit coupling to show that the off-resonant linearly polarized light can be used to tune the strength of the DMI and induce a large DMI anisotropy in the two orthogonal directions. This effect leads to the appearance of novel anisotropic phases --- an elliptic Skyrmion crystal phase and a stripe phase --- and can provide a new tool to control the stability, size, and shape of individual Skyrmions, as well as a control over the stripe phases by changing the light polarization direction. 

The predicted effects may be observed in thin films such as Co/Pt using femtosecond laser pulses \cite{Huisman2016}. The light pulses must be sufficiently long to drive the structure into a nonequilibrium state that can be considered quasistationary. The typical experimental facility with the pump fluence 2-3 mJ/cm$^2$ should be sufficient to test the theoretical results. We also expect that qualitatively the same physics persists at room temperature, helping to create a controlled set of Skyrmions that can be used to make the concept of Skyrmion racetrack memory viable \cite{Monchesky2015}.

We thank Alexey Kimel for helpful discussions. The support from the Russian Science Foundation under Project No. 17-12-01359, from the Dutch Science Foundation NWO/FOM 13PR3118, and the EU Network FP7-PEOPLE-2013-IRSES Grant No. 612624 ``InterNoM'' is gratefully acknowledged.

\supplementarystart

\centerline{\bfseries\large ONLINE SUPPLEMENTAL MATERIAL}
\vspace{6pt}
\centerline{\bfseries\large Light-induced anisotropic skyrmion and stripe phases in a Rashba ferromagnet}
\vspace{6pt}
\centerline{Dmitry Yudin, Dmitry R.\,Gulevich, and Mikhail Titov}
\begin{quote}
In this Supplemental Material we provide technical details for analytical calculations and numerical routine delivered in the main text.
\end{quote}

\section{Derivation of the effective Hamiltonian}
\label{sec:appendixa}

The model of Eq.~(\ref{hamiltonian}) of the main text is transformed by the unitary transformation (\ref{matrix1}) to the model (\ref{hamiltonian2}) which can also be written as
\be
\label{appham2}
H=\sum\limits_{n=-\infty}^\infty h_ne^{in\omega t}, \quad \mathrm{with}\quad h_n=\left[\frac{p^2}{2m}+\left(\alpha p_y+\Delta m_x\right)\sigma_z\right]\delta_{n,0}+\left[\Delta m_z\sigma_x+\left(\alpha p_x-\Delta m_y\right)\sigma_y\right]J_n(\gamma).
\e
where $J_n(\gamma)$ is the $n$-th order Bessel function of the first kind and $\gamma=2e\alpha A_0/\hbar\omega$.

If the frequency $\omega$ of a driving field is the largest energy scale in the system (the frequency $\omega$ is much larger than the bandwidth so that optical excitations of electrons are forbidden) one may replace the time-dependent problem with the time-independent Hamiltonian of an effective stationary model. The corresponding formalism of high-frequency expansion in the form of Brillouin-Wigner perturbation theory \cite{Mikami2016} has recently been developed for this class of problems. This formalism applied to the model of Eq.~(\ref{appham2}) gives the following effective model
\be
\label{appham3}
H_\mathrm{eff}=H^{(0)}+H^{(1)}+H^{(2)}+\ldots=h_0+\frac{1}{\hbar\omega}\sum\limits_{m\neq 0}\frac{h_mh_{-m}}{m}+\frac{1}{(\hbar\omega)^2}\bigg(\sum\limits_{m,n\neq0}\frac{h_mh_{n-m}h_{-n}}{mn}-\sum\limits_{m\neq0}\frac{h_mh_{-m}h_0}{m^2}\bigg),
\e
where we kept terms up to the second order in $1/\omega$. Substituting $h_n$ from Eq.~(\ref{appham2}) into the general expression of Eq.~(\ref{appham3}) we obtain the effective Hamiltonian
\begin{align}\n
H_\mathrm{eff}=\frac{p^2}{2m}+\bigg(1-2\beta\sum\limits_{n\neq0}\frac{J_n(\gamma)J_{-n}(\gamma)}{n^2}\bigg)\left(\alpha p_y+\Delta m_x\right)\sigma_z \\
\label{appham4}
+\bigg(J_0(\gamma)+\beta\sum_{\substack{m\neq n \\ m,n\neq 0}}\frac{J_m(\gamma)J_{n-m}(\gamma)J_{-n}(\gamma)}{mn}\bigg)\left(\Delta m_z\sigma_x+[\alpha p_x-\Delta m_y]\sigma_y\right),
\end{align}
where $\beta=(\Delta^2m_z^2+[\alpha\hbar k_y-\Delta m_y]^2)/(\hbar\omega)^2\ll 1$ under off-resonant radiation conditions. It is also clear that we can neglect the terms proportional to the sum of the Bessel functions in Eq.~(\ref{appham4}). Moreover, if the parameter $\gamma$ does not lay in the immediate vicinity of zeros of the Bessel function $J_0(\gamma)$ (the second line in Eq.~(\ref{appham4})) the system in question can be effectively described (in the original basis) by the Hamiltonian
\be
\label{appham5}
H_\mathrm{eff}=\frac{p^2}{2m}+(\alpha_yp_y+\Delta_xm_x)\sigma_x-(\alpha_xp_x-\Delta_ym_y)\sigma_y+\Delta_zm_z\sigma_z,
\e
which is the effective model of Eq.~(\ref{modelA}) considered in the main text. Similarly, one can show that all high order terms arising in $1/\omega$ expansion are negligible away from zeros of $J_0(\gamma)$.

In a close vicinity $\delta \gamma$ of the zeros of $J_0(\gamma)$ the high order terms are formally important. However, it can be shown that the actual value of $\delta\gamma$ is negligibly small. To provide a quantitative estimate of $\delta\gamma$ we evaluate the quantity
\be
C=\Big\vert\sum_{\substack{m\neq n \\ m,n\neq 0}}\frac{J_m(\gamma_0)J_{n-m}(\gamma_0)J_{-n}(\gamma_0)}{mn}\Big\vert=6\cdot 10^{-6},
\e
at the first zero $\gamma_0=2.4048$ of the function $J_0(\gamma)$. Thus, the effective model of Eq.~(\ref{model}) formally breaks down only in a tiny region around $\gamma_0$ that is determined by $\delta\gamma= \vert\gamma-\gamma_0\vert\sim 10^{-5}\beta$, which is well beyond our numerical resolution.

\section{Derivation of the DMI strength}\label{sec:appendixb}

In this section we present the derivation of antisymmetric exchange interaction which is linear in magnetization gradients and can be identified as the DMI. To analyze the polarization operator (\ref{polar}) we compute the quantity
\be
\label{funct}
N_i(\bb{k})=\frac{1}{2}\int\frac{d^2q}{(2\pi)^2}\sum\limits_{s,s^\prime=\pm}\frac{f(\ep^s_{\bb{k+q}})-f(\ep^{s'}_{\bb{q}})}{\ep^s_{\bb{k+q}}-\ep^{s'}_{\bb{q}}}\left(s\frac{\hat{\bb{e}}_i\,\cdot\,\bb{n}_{\bb{k+q}}}{g(\theta_{\bb{k+q}})}-s'\frac{\hat{\bb{e}}_i\,\cdot\,\bb{n}_{\bb{q}}}{g(\theta_{\bb{q}})}\right),
\e
where $\ep_{\bb{q}}^\pm=\hbar^2q^2/(2m)\pm\alpha\hbar qg(\theta_{\bb{q}})$, $g(\theta)=[\sin^2\theta+J_0^2(\gamma)\,\cos^2\theta]^{1/2}$, $f(\ep)$ is the Fermi-Dirac distribution, $\hat{\bb{e}}_x=\hat{\bb{x}}$, $\hat{\bb{e}}_y=\hat{\bb{y}}$, and $\bb{n}_{\bb{q}}=\bb{q}/\vert\bb{q}\vert$. Furthermore, one can show that
\be
\Pi_{xz}(\bb{k})=-\Pi_{zx}(\bb{k})=J_0(\gamma)\Delta_x\Delta_zN_x(\bb{k}), \quad \mathrm{and} \quad \Pi_{yz}(\bb{k})=-\Pi_{zy}(\bb{k})=\Delta_y\Delta_zN_y(\bb{k}),
\e
where we kept the notations of the main text. Taking the integral in Eq.~(\ref{funct}) we conclude that the quantities $N_i^{(1)}(\bb{k})$ in the linear order with respect to the momentum  $\bb{k}$ are given by
\be
N_x^{(1)}(\bb{k})=-\frac{1}{\pi\alpha\hbar}\int\limits_0^{2\pi}\frac{d\theta}{2\pi}\frac{k_x\cos^2\theta+k_y\sin\theta\cos\theta}{\sin^2\theta+J_0^2(\gamma)\cos^2\theta}=-\frac{k_x}{\pi\alpha\hbar}\frac{1}{|J_0(\gamma)|}\frac{1}{1+|J_0(\gamma)|},
\e
and
\be
N_y^{(1)}(\bb{k})=-\frac{1}{\pi\alpha\hbar}\int\limits_0^{2\pi}\frac{d\theta}{2\pi}\frac{k_y\sin^2\theta+k_x\sin\theta\cos\theta}{\sin^2\theta+J_0^2(\gamma)\cos^2\theta}=-\frac{k_y}{\pi\alpha\hbar}\frac{1}{1+|J_0(\gamma)|}.
\e
Thus, the second order correction $\delta\mathcal{S}[\bb{m}]$ to the bare action $\mathcal{S}_0[\bb{m}]$ reads
\be
\mathcal{S}[\bm{m}]=\mathcal{S}_0[\bb{m}]+\delta\mathcal{S}[\bb{m}]=\mathcal{S}_0[\bb{m}]+\int d^2r\left(D_x\Lambda_{xz}^{(x)}+D_y\Lambda_{yz}^{(y)}\right),
\e
which coincides with Eq.~(\ref{deltaS}) of the main text with renormalized DMI strength $D_x$, $D_y$ and Lifshitz invariants $\Lambda_{ij}^{(l)}$ defined by Eqs.~(\ref{dzyamo}) of the main text.

\section{Details of the numerical simulations}\label{sec:appendixc}

In our numerical calculations we use the lattice model given by the Eq.~(\ref{E}). 
To find the stationary states minimizing the total energy~\eqref{E} we evolve
the overdamped Landau-Lifshitz-Gilbert (LLG) equation
\be 
(1+\alpha^2)\frac{d\bb{S}_{\bb{r}}}{dt} = - \gamma\bb{S}_{\bb{r}}\times\mathbf{H}_{eff} 
-\alpha\gamma\,\bb{S}_{\bb{r}}\times\bb{S}_{\bb{r}}\times \mathbf{H}_{eff},
\label{LLG-norm}
\e
until the stationary state is reached. Here, the effective field $\mathbf{H}_{eff}$ is defined by the functional derivative of the energy~\eqref{E} over the local magnetic moment,
\be 
\mathbf{H}_{eff} = -\frac{1}{\hbar\gamma}\frac{\delta E}{\delta \bb{S}_{\bb{r}}}=H_{ext}\,\hat{\bb{z}} + J\sum_{\bb{d}=\pm\hat{\bb{x}},\hat{\bb{y}}} \bb{S}_{\bb{r}+\bb{d}}
+D_x\,(\bb{S}_{\bb{r}+\hat{\bb{x}}}-\bb{S}_{\bb{r}-\hat{\bb{x}}}) \times\hat{\bb{y}}-D_y\,(\bb{S}_{\bb{r}+\hat{\bb{y}}}-\bb{S}_{\bb{r}-\hat{\bb{y}}})\times\hat{\bb{x}}.
\e 
The stationary state corresponds to the minimum of the total energy~\eqref{E}. To ensure that a particular state is a global minimum rather than the local minimum we compare the energies of essentially different solutions obtained for each known nontrivial phase (skyrmion crystal, stripe phase or a ferromagnetic phase). To solve numerically the LLG equation~\eqref{LLG-norm} we use an explicit finite element method implemented in C and parallelized with OpenMP. Below we discuss how each of the two figures presented in Fig.~2 of the main text was obtained.

The Fig.~2a representing stability of a skyrmion is obtained as follows. We start from a seed approximate solution in the form of a skyrmion and evolve the LLG equation~\eqref{LLG-norm} for sufficiently long time to obtain the numerically exact stationary solution for an isolated skyrmion. On the second stage, the stationary skyrmion solution is distorted by adding a random noise and by varying the parameters $\gamma$ and $H_{ext}$. The solution has been accepted as a stable if, upon addition of the noise and variation of the parameters it settles down and does not decay during sufficiently long evolution time of about $\sim 3000\, \hbar/J$.

\begin{figure}
\centerline{\includegraphics[width=1.0\columnwidth]{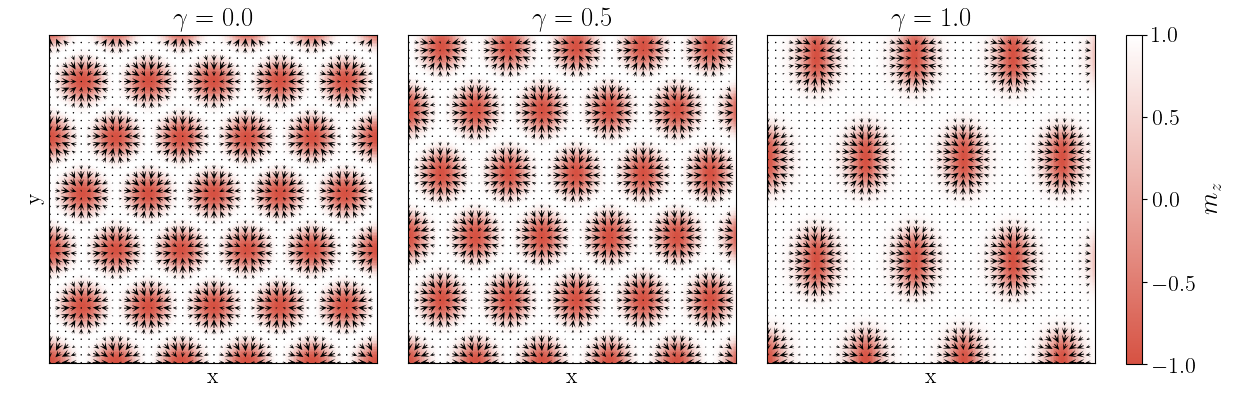}}
\caption{
Deformation of the skyrmion lattice with the parameter $\gamma$ at a fixed magnetic field $H_{ext}=0.015J$. The plots are obtained by numerical analysis of the skyrmion lattice configurations which minimize the total energy~\eqref{E}. With increasing $\gamma$ the skyrmion lattice deforms from the equilateral triangular lattice (at $\gamma=0.0$) to a isosceles triangular lattice (at $\gamma=0.5$ and $\gamma=1.0$). Square parts of a skyrmion lattice with $210\times 210$ spins are shown.  The arrows represent the in-plane components of the average magnetization in a block of $5\times 5$ spins.
}
\label{fig:lattices} 
\end{figure}

To calculate the phase diagram in Fig.~2b we proceed as follows. Configurations minimizing~\eqref{E} in a periodic rectangular domain of $N_x\times N_y$ cites are found as stationary solutions of the Eq.~\eqref{LLG-norm} obtained by evolving~\eqref{LLG-norm} from an initial seed solution for all known nontrivial phases (skyrmion crystal or stripe phase). In case of the modulated phases (skyrmion and stripe phases) the periodicity of the system for an infinite system is not known beforehand and should be found by minimizing the lattice parameters. To perform this task numerically we analyze a rectangular region of the skyrmion lattice made of $N_x\times N_y$ spins, which contains two skyrmions. The energy configuration minimizing the energy density $E/N_xN_y$ is then found by the coordinate descent method in the configuration space ($N_x$,$N_y$). The minimal energy configurations obtained for different values of the parameter $\gamma$ and for a fixed magnetic field $H_{ext}=0.015J$ are shown in Fig.~\ref{fig:lattices} where a part of the skyrmion lattice with $210\times 210$ spins is shown. As seen from the Figure, the increase of $\gamma$ deforms the skyrmion configuration from the equilateral triangular lattice (at $\gamma=0.0$) to a isosceles triangular lattice (at $\gamma=0.5$ and $\gamma=1.0$). Similar procedure is implemented for the stripe crystal. In this case the problem is simplified since the optimization is required only along one direction (due to the tendency of stripes to align along the ``easy'' direction defined by the polarization of the applied radiation). The energies of the stationary solutions obtained are, then, compared to the minimal energy identified as a ground state.

\begin{thebibliography}{99}
\bibitem{footnote} The frequency $\omega$ is assumed to be larger than the bandwidth. Naturally it is also assumed that the condition $\omega\tau \gg 1$, where $\tau$ is a characteristic electron relaxation time, is very well fulfilled.
\end{thebibliography}
\end{document}